# Detection of Syntactic Aspect Interaction in UML State Diagrams Using Critical Pair Analysis in Graph Transformation

Zaid Altahat[1], Tzilla Elrad[2], Luay Tahat[3] and Nada Almasri[3]

[1] GE Healthcare
Chicago, IL. USA

[2] Illinois Institute of Technology (IIT)
Chicago, IL. USA

[3] MIS Department, Gulf University for Science and Technology
Mishref, Kuwait

**Abstract**
Aspect Oriented Modeling separates crosscutting concerns by defining Aspects and composition mechanisms at the model level. Composition of multiple Aspects will most likely result in more than one Aspect matching the same join points. Consequently, Aspects do not always interact in a predictable manner when woven together. Intended interaction among aspects is designed by the system designer. Unintended interaction (or interference) must be automatically managed. When the woven aspect demonstrates a behavior that is different than its autonomous behavior, then this is a potential interference. Interference has been recently reported in Aspect Oriented Software Development (AOSD) by the industry. Leaving this problem unsolved may result in erratic software behavior and will hinder the adaptation of AOSD by the industry. This identified problem is similar to a phenomenon that exists in graph transformation systems where multiple Graph Transformation rules share some conflicting elements, it is referred to as Critical Pair Analysis and it provides an algebraic-based mechanism to detect and analyze the interaction of the rules. In this paper we propose a framework to detect unintended Aspect interaction at the model level. The proposed framework transforms Aspects modeled in UML State Diagram to Graph Transformation Rules, and then it applies Critical Pair Analysis to detect unintended interactions among aspects. This will enable developers to specify only the order of precedence for intended interaction among aspects without the need to manually investigate unintended interactions for the combinations of every Aspect to every other Aspect in the system. The proposed interaction detection solution is automated, modular, and independent of the base model; which adds the advantage of not having to re-evaluate the interaction each time the base model changes.

## 1. Introduction

Software modules are added to other software modules and other components in an incremental way to build software products. This process will most probably result in interaction among the software modules. Most software modules have several complex interactions with other software module through their life cycle. Interference among independently built software modules plays a critical role in undermining the stability of the software product under development [10, 16, 20, 40, 41].

The modular approach of software development that is followed by Aspect Oriented Software Development AOSD [42] makes susceptible to the interference problem among the independently built aspects. The behavior of the software may be unpredictable due to the interference problem. In order to have a predictable behavior by the product, the interference must be eliminated. The elimination of the interference requires a practical approach of detecting all possible interfering aspects. AOSD modularizes concerns that crosscut with other concerns into aspects, which are later woven to the rest of the base system's program (or model). Woven aspects do not usually work in isolation; aspects collaborate to deliver a service. Desired cooperation between aspects is manually designed into the system. Undesired, or unplanned, cooperation on the other hand must be precisely defined using an automated approach. The interference (or unintentional interaction) among the software modules may result in a random behavior of the software product that is dependent on the weaving order. We believe that the interference that might take place among aspects must be automatically and efficiently detected.

This interaction was investigated earlier on the telephony systems and referred to as Feature

Interaction [3, 6, 10, 13]. Different mechanisms [6, 10, 13, 34] were proposed to handle the FI problem. Aspect Oriented Software Development (AOSD) [35] builds software systems by composing crosscutting concerns in a similar approach to the features in the telephony systems. This leads to the Aspect Interaction (AI) problem that is very similar to the FI problem. The AI is not necessarily harmful [25]. But the term AI usually refers to the unintended interaction. If the interaction is planned, order precedence needs to be defined [20]. If a dependency between two Aspects is not planned, then unless an AI detection mechanism is used, the dependency might slip undetected with potential harm to the system. The Motorola WEAVR [17] has reported the AI problem in the Telecomm industry, where precedence is defined for interacting Aspects [20]. Graph Transformation (GT) systems have developed a mechanism to detect conflicts among GT rules [38].

In this paper, we propose a graph-based framework to detect unintended interaction among Aspects in UML State Diagrams. Our approach uses Critical Pair Analysis (CPA) which is a technique originally used in telephony systems to detect features interactions. In our approach we transform Aspects defined in UML state diagrams to GT rules, and then we apply CPA to automatically detect interacting aspects.

This paper is organized as follows; Section 2 describes the different types of Aspect interactions. Section 3 presents the GT systems and CPA. Section 4 presents our proposed framework with an example of an ATM modeled as State Diagram. In the case study presented in Section 5, multiple Aspects are defined and their interactions are analyzed and classified using the generated CPA report. Related work is discussed in section 6. Conclusion and future work are discussed in section 7.

## 2. Aspect Interaction

With the use of AOSD to manage separation of concerns, AI is an inevitable issue. AI takes place when multiple Aspects share conflicting elements in their pointcuts or advices. Multiple aspects are said to be independent if the order of applying aspects result in the same model. Two models are considered syntactically the same if there is a bijective mapping between the two models. That is given two models M1 and M2, for each element in M1 there is one element in M2 with the same properties. There is also the same reverse mapping from M2 to M1.

Interaction among Aspects exists in the form of either dependency or conflict. This kind of AI is referred to as Aspect-to-Aspect interaction [20]. Even non-conflicting and independent Aspects might have unintended impact on the structure of the base model; this kind of interaction is referred to as Aspect-Base interaction. Aspects may also have unintended impact on the behavior of the base model, this kind of interaction is referred to as *semantical* interaction [5]. The next 4 definitions will shed light on the different types of interactions. Let:

$M_1$ = The result of applying Aspect A1 to the Base Model (BM).
$M_2$ = The result of applying Aspect A2 to the BM.
$M_{12}$ = The result of applying A1 *then* A2 to the BM.
$M_{21}$ = The result of applying A2 *then* A1 to the BM.
**Definition 1:** Two Aspects do not have interaction between them iff $M_{12} = M_{21}$.
**Definition 2:** A *dependency* exists between two Aspects if ($M_{12} = M_2$) or ($M_{21} = M_1$)
**Definition 3:** A *conflict* exists between two aspects if ($M_{12} = M_1$) or ($M_{21} = M_2$)

Definition 1 states that, regardless of the order of applying Aspects, the output model is the same. This is only possible if the application of one Aspect does not alter the applicability of the other Aspect. If $M_{12} \neq M_{21}$, then AI exists between A1 and A2 in the form of either dependency (definition 2) or conflict (definition 3). If ($M_{12} \neq M_{21}$ And $M_{12}=M_2$) then A1 depends on A2, or if ($M_{12} \neq M_{21}$ And $M_{21}=M_1$) then A2 depends on A1. A conflict is defined as either ($M_{12} \neq M_{21}$ And $M_{12}=M_1$) or ($M_{12} \neq M_{21}$ And $M_{21}=M_2$), which means A1 disables A2, or A2 disables A1, respectively.

## 3. Graph Transformation and Critical Pair Analysis

This section describes Graph Transformation (GT) and Critical Pair Analysis (CPA) which are used in telephony systems to detect features interaction (FT). A graph transformation applies a GT rule ($P = L, R$) to a host graph G; where $P$ is a production, $L$ is the left hand side (LHS) graph, and $R$ is the right hand side (RHS) graph. $P$ may also have a set of Negative Application Conditions (NAC), which are elements that may not exist for a rule to apply. A GT rule replaces graph $L$ with $R$ in host graph G.

Figure 1 shows a simple graph, referred to as host graph, (left), a GT rule (middle) with its LHS ($L$) and its RHS ($R$) components, and the generated graph (right). Host graph is searched for a graph morphism of ***L***, referred to as a *match*. A graph morphism between two graphs, G and H, is a bijective mapping (ψ) between the vertices of G and the vertices of H, such that two vertices u and v are adjacent in G iff their mapping vertices ψ(u) and ψ(v) are adjacent in H. If a match is found, the graph ***R*** is applied to the host graph. GT works as follows:
- Elements in *L* and in *R* are preserved in the generated graph.
- Elements in *L* but not in *R* are deleted from the generated graph.
- Elements in *R* but not in *L* are created in the generated graph.

Figure 1-(middle) shows a NAC edge between the states '**b**' and '**c**' in the LHS, marked with '**X**'. The NAC will be used in transforming some of the pointcut constructs, such as 'XOR'. With out the NAC edge in Figure 1, the matching mechanism will only check for the existence of vertices '**b**' and '**c**' without checking the absence, or presence, of an edge between '**b**' and '**c**'. These requirements for morphism come from L. According to **R** edge '**e4**' is created and vertex '**d**' is removed. Generated graph is presented in Figure 1 (right).

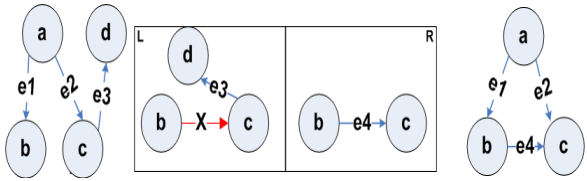

Figure 1. A graph transformation rule on a directed labeled host graph.

GT rules are used to apply changes to a host graph in order to integrate a new feature. Two GT rules that overlap are said to be critical pair. Critical Pair Analysis (CPA) is used to detect conflicts and dependencies in GT Systems. Two rules are in conflict if one rule disables another rule. On the other hand, two rules are dependent if one rule enables the other. Critical Pair Analysis (CPA) [23] is "a pair of transformations both starting at a common graph G such that both transformations are in conflict, and graph G is minimal according to the rules applied." [32]. That is the GT rules **P1** and **P2** form a critical pair if both, **P1** and **P2**, can be applied to the same minimal graph G. But applying **P1** will prohibits the application of **P2** and/or vice versa. Certain tools, such as Attributed Graph Grammar (AGG) [1], provide graph transformations and CPA. An attributed graph allows the definition of attributes on graph elements. CPA and NAC [38] are combined to detect conflicts in GT systems.

## 4. Aspect Interaction Detection Framework

Aspect Oriented Modeling (AOM) [19] follows an approach similar to the GT systems by querying and adapting base model elements. If a mapping is created between GT rules and Aspects in AOM, then the CPA technique can be used to analyze AI to detect any conflicts and dependencies among Aspects

Since applying Aspects to a base model involves matching and modifying elements in the base model similar to those of the GT systems, our approach uses CPA to detect syntactic interaction among aspects in the UML State Diagrams. UML State Diagrams are increasingly used in modeling wide range of embedded devices, from small gadgets to Telecom Systems [17].

The proposed framework detects potential conflicts and dependencies among Aspects without the need to check the base model for the pointcuts applicability. It achieves this by inspecting all combinations of the pointcuts and advices of all Aspects in a pair-wise manner. This approach will report potential interaction among all Aspects, even if some Aspects might not have a match and regardless of the base model. The advantage of this approach is to avoid regenerating the AI report each time the base model changes.

The framework has a tool which automatically transforms Aspects modeled in UML state diagram to GT rules. Additionally, it uses AGG tool [1] to generate interaction report based on CPA analysis applied on the provided GT rules. **Figure 2 Framework Structure** shows the architecture of the framework where detecting aspect interaction is done in two phases. In Phase one, Aspects are transformed to GT rules automatically using Aspect-to-GT tool which was developed. In phase two, the GT rules obtained from phase 1 are fed to AGG tool which applies CPA analysis and generates the interaction report. The generated report has two matrixes, one showing the minimal conflicts between all pairs of Aspects, and the second matrix shows the minimal dependencies between all pairs of Aspects.

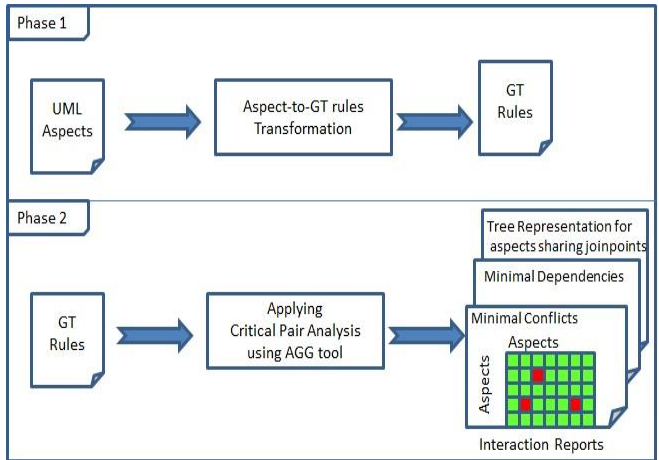

Figure 2 Framework Structure

In section 4.1 we explore the main characteristics of our proposed solutions. In section 4.2, we demonstrate how Aspects modeled in UML state diagram are transformed to GT rules, and in section 4.3 we explain how AGG tool is used to generate CPA interaction reports.

## 4.1. Characteristics of the Framework

Our proposed solution to the aspect interference problem provides an interference detection solution that is 1) Automatic, does not require the user's intervention, 2) Modular, analysis is done independently of the base system, 3) efficient, the performance of the approach is reasonable, and 4) Practical, does not require extra level of expertise. Aside from the characteristics of our proposed solution, using our solution may eliminate the extremely laborious and near impossible manual process of identifying interfering aspects and increase the confidence in AOSD and bring it one step closer to the industry.

Automation and modularity are essential to any interference detection approach. An approach is modular if the detection analysis is performed on the aspects and the base model separately. Using the proposed approach, users will be able to identify interacting Aspects independently of the base model. This way users need to define only order precedence for identified Aspects once, when they are defined, not each times the model changes. So changes to the base model will not affect the AI, only changing Aspects will require a re-run of the analysis.

Without Automation for Aspect interactions, Aspects designers would have to specify order precedence for all Aspects in the system. This requires large efforts parts of which are useless and wasted. Aspects' designers may still be interested in specifying order for certain Aspects, but they do not have to specify it for all Aspects.

Currently our proposed framework studies the Aspect-to-Aspect interaction; the other types of interactions discussed in section 2 are planned for future work.

## 4.2. Transformation of Pointcuts to Graph Transformation Rules

In this section we present an example that will demonstrate how Aspects with pointcuts consisting of composite state and compound transitions are transformed to some GT rules. The example consists of an ATM machine described by the UML State Diagram presented in Figure 3. The ATM lacks the behavior to diagnose and early terminate the ATM machine. The behavior is added to the base ATM model by the 'diagnostic' concern, presented in section 4.2, that has 4 Aspects.

Figure 3 presents the UML State Diagram for a bank ATM. Since the *Active* state is composite non-orthogonal, only state '*validating*' will have the incoming transition '*card_in*'. The *Maintenance* state is orthogonal, so both states '*testing*' and '*waiting*' will receive the incoming transitions '*maintain*'.

In order to make it easier on the reader to follow, we numbered each state in Figure 3 and used the numbers in the generated GT rules. Vertices whose names are separated by a '|', for instance the vertex '**1|4**', represent substates in the composite orthogonal state *Maintenance*. Digits to the left of '|' come from the top region, and digits to the right come from the bottom region. When the state '*Maintenance*' becomes active, states '*testing*' *(1)* and '*waiting*' *(4)* become active.

The following 4 Figures, 4 through 7, present the 4 Aspects which are part of the concern 'diagnostic' that will add the behavior to diagnose and early-terminate the ATM. Figure 4-(a) presents the first Aspect A1. Elements marked with '**E**' are exposed and passed to the weaver to adapt. Also to simplify presentation of the GT rules, if an element is presented in the LHS but not in the RHS, it does not mean that the element is deleted, they are just not shown for simplicity.

The Aspect in Figure 4-a is transformed to three different GT rules shown in Figure 4-b. The three vertices (2|6), (2|5), and (2|4) represent the different states the composite state 'Maintenance' might be in while in state 'self_diagnostic' (2). The RHS of the GT rules in Figure 4-b represent the creation of the

edge 'diagnostic' between the states 'validating' (9) and 'self_diagnostic' (2).

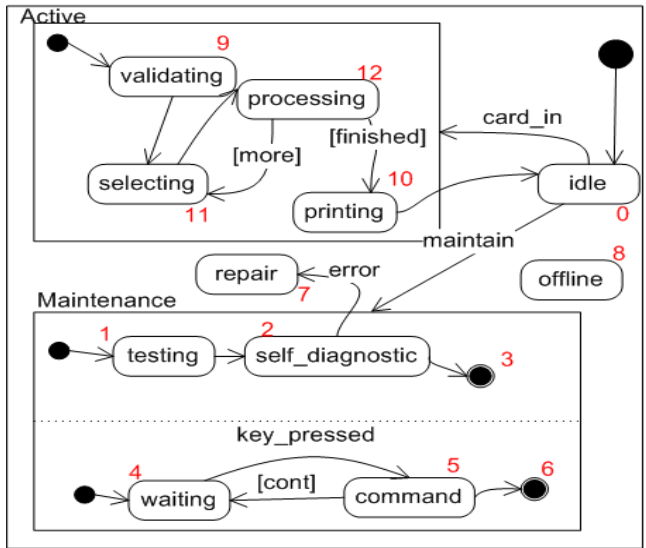

Figure 3 UML State Diagram of an ATM

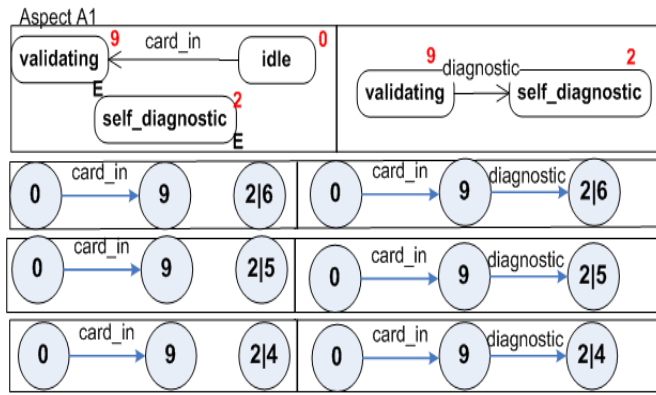

Figure 4  a-(top) Aspect 1, b-(bottom) GT rules

Figure 5 presents the second Aspect A2. The advice of the Aspect creates the edge 'eject' to the sequential state 'Active'. Every substate in the state 'Active' will have an incoming edge labeled 'eject' from the state 'idle'. The vertices 9,10,11, and 12 of the generated GT rules in Figure 5-b represent the substates of state 'Active'. Note, to simplify the presentation of the GT rules, the RHSs do not show the edge 'diagnostic' and the vertices (2|6), (2|5), and (2|4) which are preserved in the host graph.

Figure 6 presents the third Aspect A3. The Aspects creates the *fork* transition 'diagnostic' which forks to the two sub states, 'self-diagnostic' and the final state of the bottom region in the composite state 'Maintenance'. These results in one GT rule are shown in Figure 6-b.

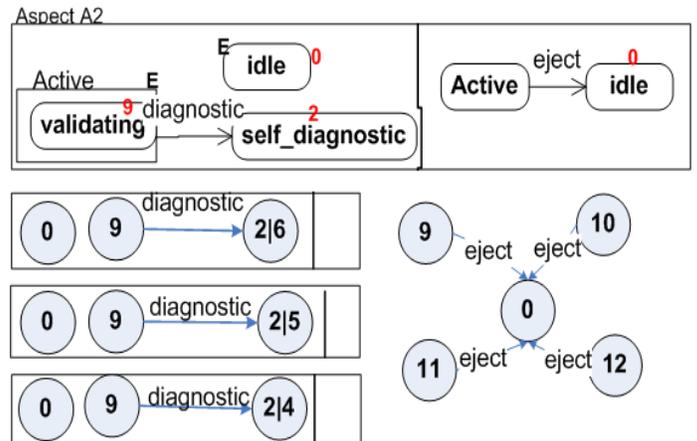

Figure 5 a-(top) Aspect 2, b-(bottom) GT rules

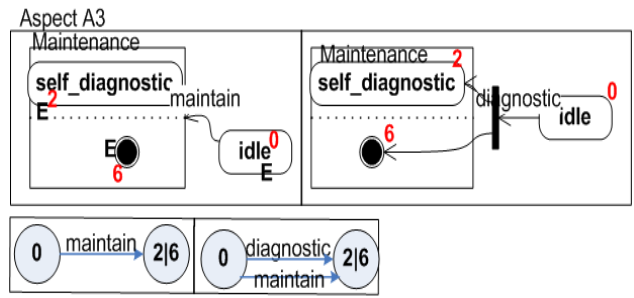

Figure 6 a-(top) Aspect 3, b-(bottom) GT rules

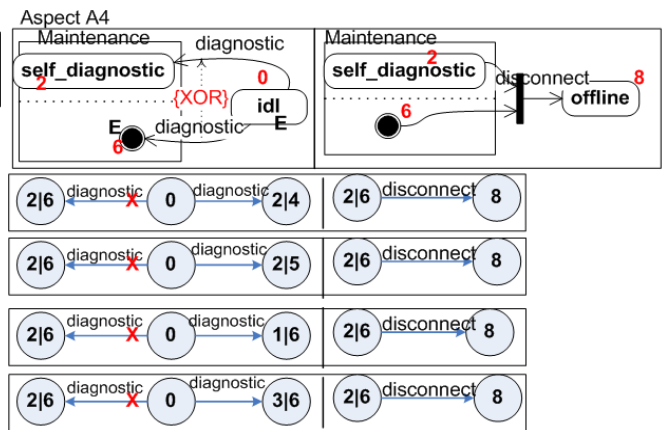

Figure 7 a-(top) Aspect 4, b-(bottom) GT rules

Figure 7 presents the fourth Aspect A4. The pointcut will match the edge 'diagnostic' from the state 'idle' to the state 'self_diagnostic', or from the state 'idle' to the state final state of the bottom region, but not from both.  The NAC is used to transform the 'XOR' element. First two of the 4 GT rules in Figure 7-b present the GT rules that match the transition 'diagnostic', and the states 'idle' and

'self_diagnostic'. At the same time it does not allow the same transition between the states 'idle' and the *final* substate. The bottom 2 GT rules show the opposite.

### 4.3. Generating CPA reports

The GT rules of the four Aspects presented in Figures 4-b, 5-b, 6-b, and 7-b are fed to the AGG to generate the CPA report. There are a total of 11 GT rules. To trace the GT rules back to the Aspects, each of the GT rule's name consist of two parts separated by a hyphen. For example the GT rule "A1-R1" represents the first GT rule (R1) of the Aspect A1, presented in Figure 4. The GT set "A1-*" refers to all the GT rules in A1. Any pair of rules in the GT set "A1-*" that is in conflict or dependency with any GT rule in other GT sets, will cause the Aspect A1 to be in the same conflict or dependency as its rule. For instance there is a conflict, Figure 8 (top), between the rules A3-R1 and A4-R1, which causes the Aspects A3, and A4 to be in conflict. This is because the pointcut of A4 doesn't allow for the transitions created by A3 for its pointcut to have a match. Also by inspecting Figure 8 bottom, we can see that A2 depends on A1 for its pointcut to find a match. One thing to mention is that any conflict or dependency within the same GT set is irrelevant and ignored. Note, due to space, Figure 8 shows only part of the report for the interacting Aspects.

## 5. Case Study: POTS

In this section, we applied our solution for the detection of aspect interference to the POTS phone system [Kor00] which consists of nearly 40 aspects. Detection of interference is achieved by composing pairs of all aspects, so we have 1600 pairs of aspects. For the 1600 pairs we generate:
- A Tree representation of all the aspects that share a joinpoint.
- A matrix of the conflicts among all aspects.
- A matrix of the dependency among all aspects.

Without our solution to the interference problem, the aspect designer has to manually inspect the 1600 pairs and decide for possible interference among the aspects.
The POTS system consists of the base model and a set of features. The base model is modeled in UML state diagrams. The features were also transformed into aspects and modeled in UML state diagrams.

Two types of interference were detected, Dependency among aspects, and Conflicts among aspects.

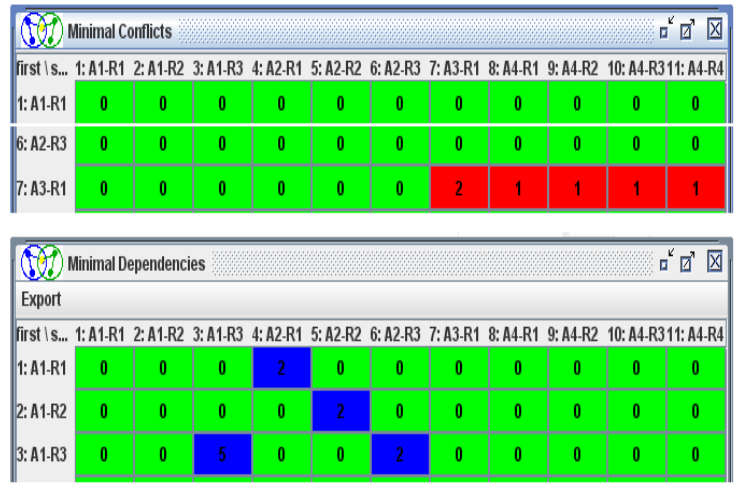

Figure 8 The CPA report of the ATM state machine

### 5.1. The Basic Model

Figures 9 and 10 (in Appendix) present two representations of the same basic call model, UML and graph, respectively. The name of all the states visited during the process of making a phone call start with "o ", for originating, to distinguish them from the receiving phone call states that start with "t ", for terminating.

### 5.2. The Concerns and their Aspects

The phone system has 10 new concerns defined in 40 aspects. See Altahat [43] for a list of all the aspects. The 10 concerns are typical phone system added features such as call forwarding, call waiting, three way calling, reverse billing, split billing, and few others.

### 5.3. Results for Aspects with Dependency

Figure 11 (in Appendix) presents the dependency results for the POTS phone system with 40 aspects. The green pairs do not have dependency among them. The shade (blue) pairs on the other hand do have dependency among them and require the designer intervention to at least define the weaving order for these pairs.

We will pick one pair with dependency between them. The aspect CW-S-A7 (Call Waiting - Subscribers), presented in Figure 12 (see Appendix) depends on the aspect CW-S-A5, presented in Figure 13 (see Appendix). The dependency is due to the state CW3:cw hold. The pointcut of the aspect CW-S-

A7 has the state CW3:cw hold. But the state CW3:cw hold does not exist in the base model and only exists after running the aspect CW-S-A5, which creates the state CW3:cw hold.

When inspecting the dependency results in Figure 11 (Appendix), we notice that certain columns (or aspects) are almost entirely shaded (blue). Such as the aspect 1:CFB-S shown in Figure 14 and the aspect 8:RB-S-A1 shown in Figure 15 (Appendix). This means that the aspects has dependency with almost every other aspect in the system.

By inspecting the aspect (1:CFB-S) we notice that the aspect's pointcut matches all states except for the state "idle". Which means that any aspect that creates new states, except for the "idle" state, will have a dependency with the aspect 1:CFB-S. Similarly, the pointcut of the aspect 8:RB-S-A1 matches that states (o o_ hook, o dial tone, o dialed, o busy, o wait for answer, o connected, o wait for onhook, t connected, and t wait for onhook). So any aspect that might create any of these states will have a dependency with the aspect 8:RB-S-A1.

### 5.4. The Results for Aspects with Conflicts

A conflict between two aspects exists if the application of one aspect prevents the application of the other aspect. Figure 16 (Appendix) presents the results for the POTS phone system with 40 aspects. The green pairs do not have conflict among them. Meaning running them in any order will generate the same model and consequently does not require the designer intervention. The shaded (red) pairs on the other hand do have conflict among them and require the designer intervention to at least define the weaving order for these pairs.

## 6. Related Work

Several researches has been conducted by the Aspect-Oriented Software Development (AOSD) community in order to reduce complexity and increase reuse of the software by providing modularization of concerns that tend to crosscut. In particular, several researches have been conducted to deal with the problem of Aspect Interaction at different phases of the software development mainly at the requirements level [26, 27, 28, 29]. Khan et al [27] proposed multidimensional concern slicing approach that simplifies identifying requirements dependencies and impact of requirements changes. Dependency graphs for each concern slice are constructed from semi-formal dependency equations, which assist in analyzing the consequential change impact on the requirements. The early determination of requirements dependencies may reduce the undesirable ripple effect propagation. Magno et all [28] studied interactions between concerns and proposed a classification that takes into account the type of concern being studied. Decisional concerns are ranked using a systematic process. Based on this, the orders of concerns' composition were derived. Their approach provides a better understanding of the interactions between all the elements of a system.

S. Ciraci et al [44] proposed graph-base model checking of AOM for Aspect interference. Their approach relies on transforming both the base model and Aspects to Design Configuration Modeling Language (DCML) which is a language proposed by the authors. They run execution simulations to check for Aspect interference. Their work mainly focuses on UML class diagrams and UML sequence diagrams.

Bar-On et al [29] proposed a method that supports the identification of functional requirements that crosscut other functional requirements to generate the derived or modified requirements. To identify crosscutting requirements, they manually used match actions used by requirements and the system modes and states related to these requirements. Their method is based on the observation that, when the same action is used by two requirements it indicates that one of the requirements may crosscut the other.

Shaker and Peters [31] proposed a process for detecting undesirable concern interactions in AO systems at the design phase of the software development process. They describe a statechart weaving language for specifying and later verifying the weaving into an UML model at design stage. Kienzle et al [8] studied Aspect interaction for Aspect-Oriented Programming environment such as AspectJ. They have defined an aspect based on the services it requires from other aspects and on the services it removes. They also established a set of composition rules to solve inter-aspect dependencies. The Aspect Interaction Aspect-Oriented Programming is also classified into different types in [36, 37].

When AOSD was first introduced by Douence et al [14], they studied the interaction problem and proposed a framework for detecting aspect interactions at the language level for AOP. Douence et al considered among the first to look at this problem. Order Precedence for the Aspect-to-Aspect interference of models in the Motorola WEAVR [17] was proposed by Zhang et al [20]. They define three precedence relations as *follows*, *hidden_by*, and

*depdendent_on*. Their intent was not to detect interaction, but rather to define precedence relations for interacting Aspects. Mostefaoui et al [5] proposed semantic conflicts between aspects and base model. They translated models to Alloy [2] to be formally verified. Their approach is for semantic verification of aspects and base model interaction. For each aspect they define constrains, pre and post conditions, that will be verified using Alloy at the weaving time.

Bakre and Elrad [33] used Live Sequence Charts to detect AI at the Joinpoint in the form of use-case scenarios. They proposed the Aspect Interaction Charts that build on top of the Live Sequence Charts in order to capture the interactions among various aspects at joinpoints. The Aspect Interaction Charts has the ability to capture aspect interactions at a joinpoint in a common specification in the form of use-case scenarios, and the ability to execute these scenarios while non-invasively manipulating the interactions among the various aspects. They used the tools that come with the Live Sequence Charts language, i.e. the Play Engine to model, view and manipulate aspect interactions at joinpoints.

Havinga et al proposed a graph-based approach [18] to detect composition conflicts due to weaving multiple aspects in AspectJ [22]. They model the structure of programs as graphs and the model introductions as graph transformation rules. They defined explicit rules to describe when composition conflicts related to introductions occur. A prototype tool has been built to detect and visualizes the occurrence of such conflicts in AspectJ programs. The graph-based models are generated automatically from the source code of Java programs and AspectJ introductions. Their approach did not make strong assumptions about either the aspect or base language; it has been designed to be applicable to other AOP languages.

GROOVE [21] is a project centered around the use of simple graphs for modeling the design-time, compile-time, and run-time structure of object-oriented systems, and graph transformations as a basis for model transformation and operational semantics. In their approach they detect pre-defined language violations, such as multiple conflicting method definitions, and cyclic inheritance. The essence of their work is to verify predefined rules in AspectJ, contrary to our approach, which is to detect conflicts among aspects.

Nagy et al [7] proposed a method of analysis of aspect interaction in AOP that was applied to AspectJ. They provide a solution that is constraint-based and declarative for interacting aspects. Their work does not discuss mechanisms for detecting interaction among aspects. Our work concentrates on detecting interaction, dependency and conflicts, of aspects. A mechanism for semantic aspect interaction in Composition Filters for AOP is proposed by Durr et al [11]. They provided a mechanism similar to the mechanisms for detecting deadlock in a computer system. Based on the semantics of the added advices, their approach tries to order aspects in a harmless way.

Mehner et al [9] proposed an approach for analyzing interactions and potential inconsistencies at the level of requirements modeling using variant of UML. Critical Pair Analysis is used to analyze aspect interaction in UML class diagrams. Model transformations are expressed as pre and post conditions that are used in defining graph transformations rules. Pre and post conditions are derived from activity diagrams. In their approach classes and associations among classes are tracked using Attributed Graph Grammar to analyze their interaction. The analysis is performed with the graph transformation tool Attributed Graph Grammar. The automatically analyzed conflicts and dependencies also serve as an additional view that helps in better understanding the potential behavior of the composed system.

Jayaraman et al [12] used Critical Pair Analysis to detect feature interaction in Software Production Lines. Their work presented a graph-based Modeling Aspects using a Transformation Approach to specify how features, modeled in UML, relate to each other. Our framework is also graph based but for UML State Diagrams, in particular composite states and compound transitions and their transformation to GT rules. However, our framework uses CPA technique to detect Aspect-to-Aspect Interaction.

## 7. Conclusion and Future work

In this paper, we demonstrated how to detect AI in UML State Diagrams. The proposed framework uses Critical Pair Analysis in the GT Systems to detect the interaction, CPA is provided by AGG. The framework has a complexity of $O(n^2)$, where "n" is number of Aspects; but the AI detection for a pair of Aspects needs to be done only once in the system's lifetime. Hence the introduction of a new Aspect to the system will result in (n) pairs between the new and existing Aspects; AI among existing Aspects doesn't need to be reevaluated. Consequently, only a

O(n) is needed for the introduction of a new Aspect. The proposed approach is modular (independent of the base model). This adds a huge advantage in large industrial system.

In order to use CPA, Aspects are transformed to GT rules. KerMeta was used to execute all the model transformations. As seen in section 4, users do not have to define order precedence for all possible combinations of Aspects. Instead user is required only to define order between the Aspect A1 and A2 and precedence between A3 and A4.

However, the proposed framework does not support pattern matching in defining pointcuts, similar to those supported by AspectJ. This is due to the limitation enforced by AGG. There are also other mechanisms that are more expressive, such as Join Point Designation Diagram (JPDD) [4, 15] and the State Machine Joinpoint Model [16] used in the WEAVR [17]. Such mechanisms will result in different GT rules when integrated into our framework. In future work we plan on adding support for JPDD in our framework. As seen in section 4, traceability between the GT rules and Aspects was done manually by using the Aspect#-Rule# naming convention. In large-scale production an automatic traceability is needed which will automatically decide which Aspects are in conflict or dependency without having to report the triggering GT rules.

## Appendix : POTS Case Study Figures

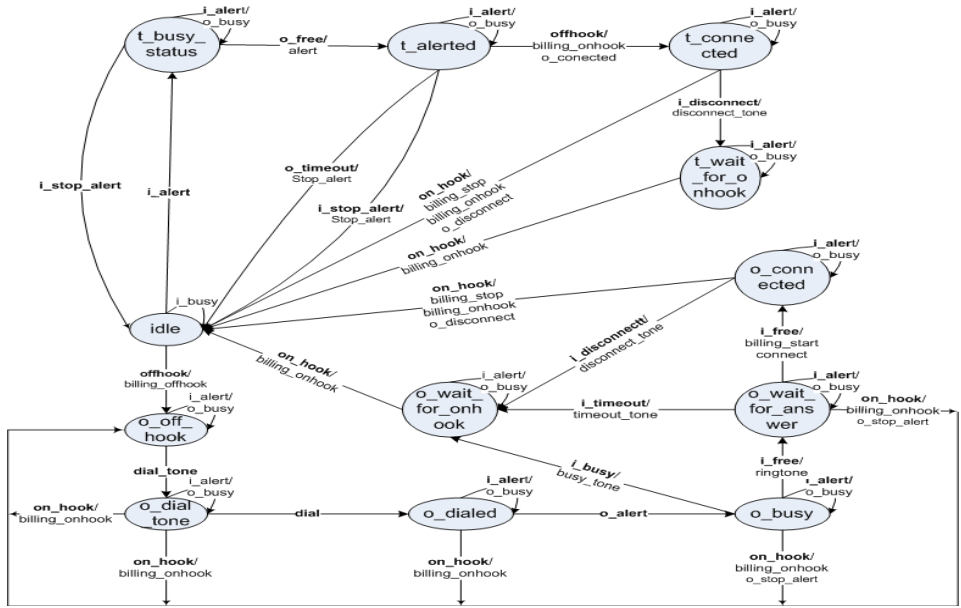

Figure 9 The Basic Call Model in UML

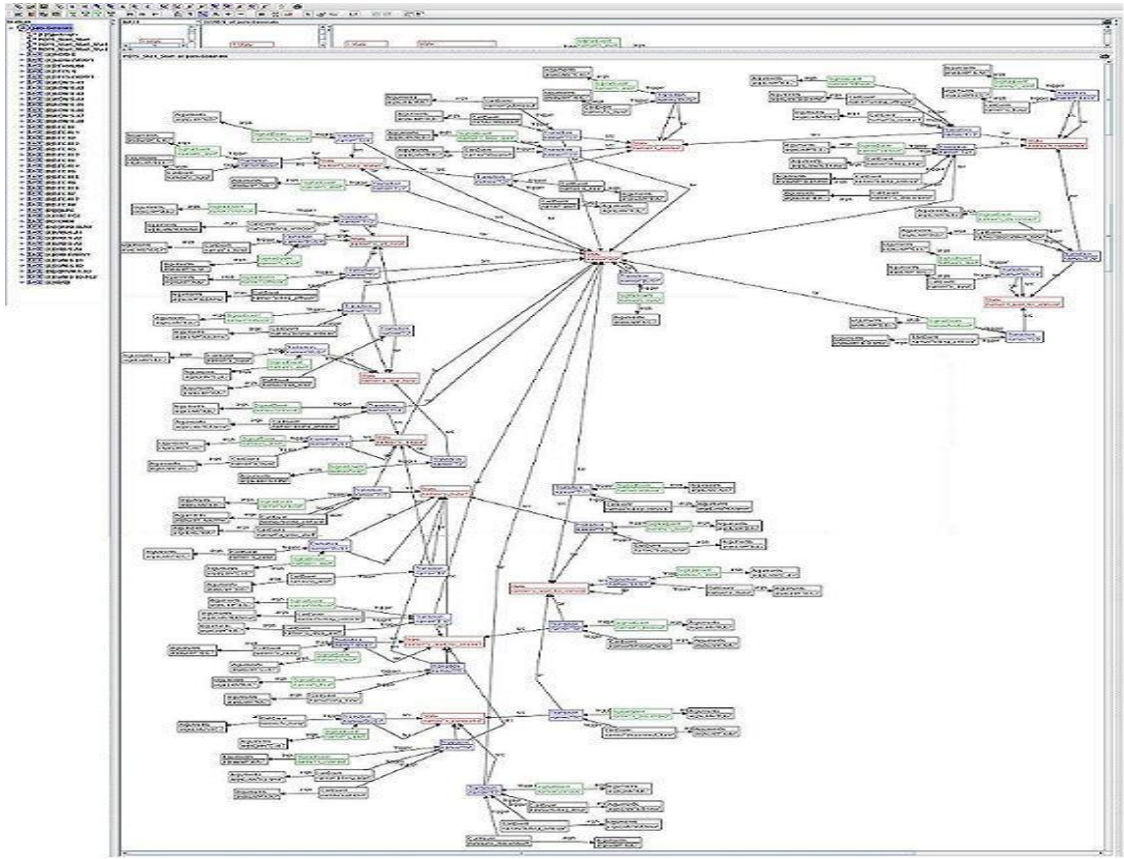

Figure 10 The Basic Call Model in Graph

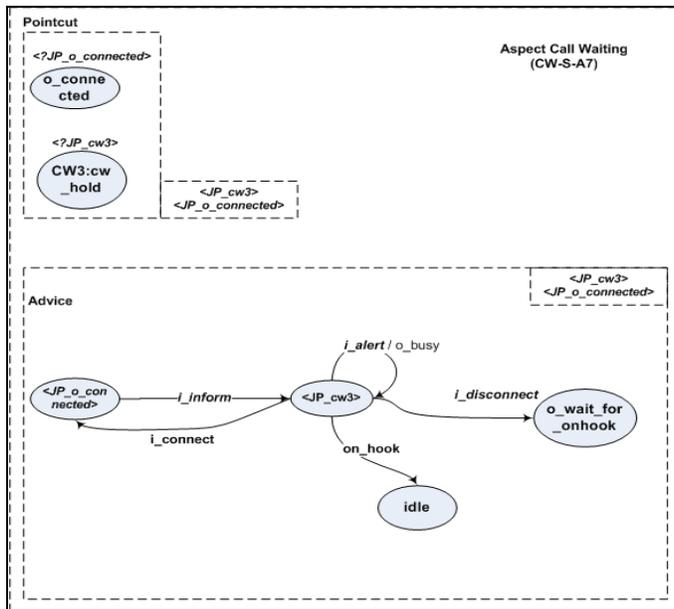

Figure 11 The Aspect Call Waiting for Subscriber (A7)

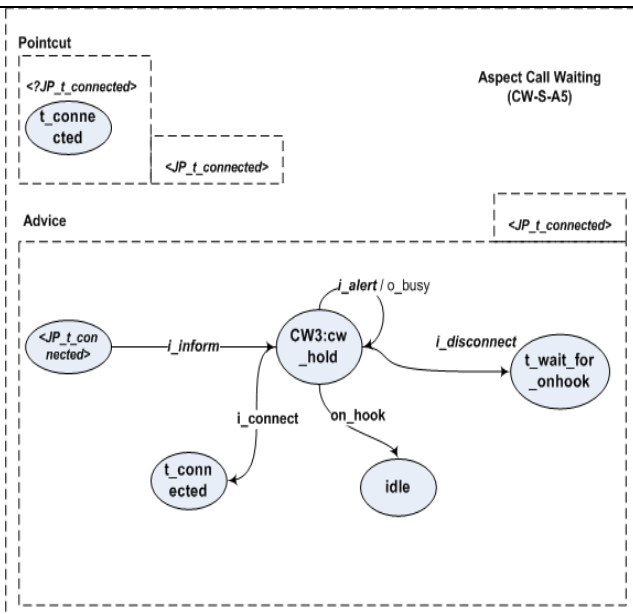

Figure 12 The Aspect Call Waiting for Subscriber (A5)

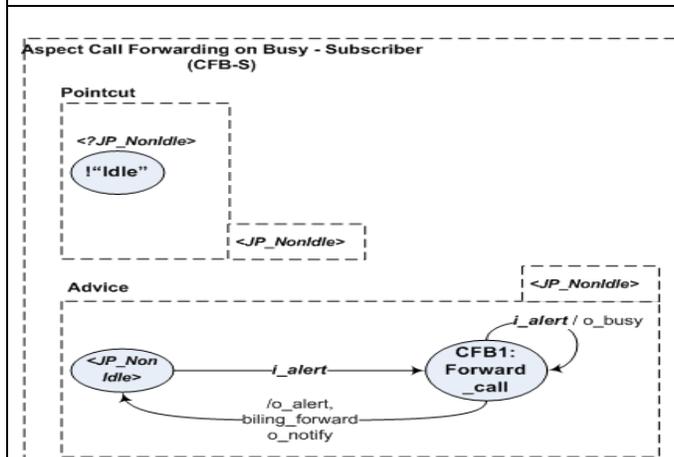

Figure 13 The Aspect Call Frowarding on Busy for Subscriber

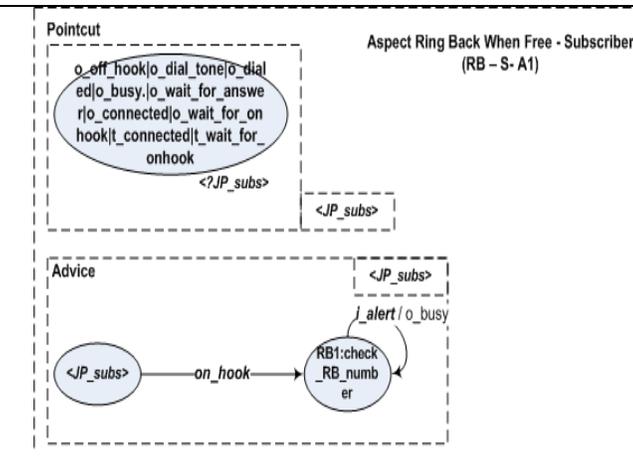

Figure 14 The Aspect Ring Back When Free: Subscribers (A1)

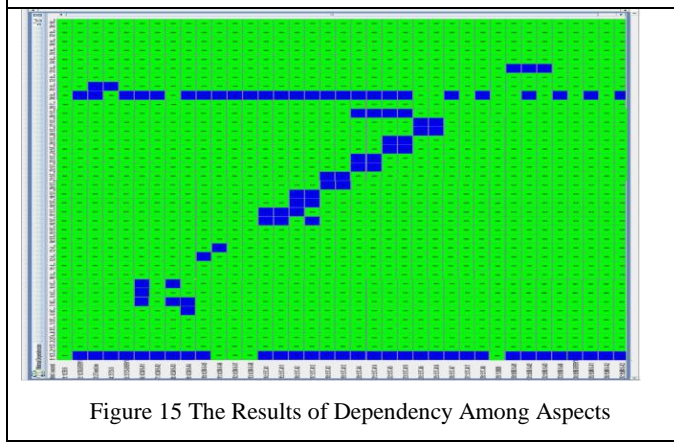

Figure 15 The Results of Dependency Among Aspects

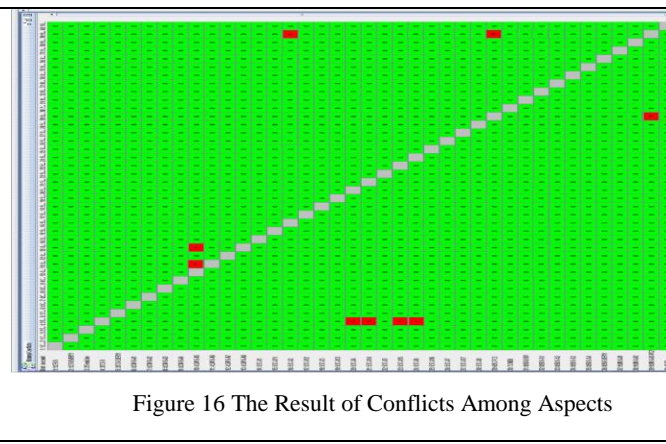

Figure 16 The Result of Conflicts Among Aspects